\documentclass{article}

\usepackage{arxiv}

\usepackage[utf8]{inputenc} % allow utf-8 input
\usepackage[T1]{fontenc}    % use 8-bit T1 fonts
\usepackage{hyperref}       % hyperlinks
\usepackage{url}            % simple URL typesetting
\usepackage{booktabs}       % professional-quality tables
\usepackage{amsfonts}       % blackboard math symbols
\usepackage{nicefrac}       % compact symbols for 1/2, etc.
\usepackage{microtype}      % microtypography
\usepackage{lipsum}		% Can be removed after putting your text content
\usepackage{graphicx}
\usepackage{natbib}
\usepackage{doi}
\usepackage{multirow}
\setcitestyle{numbers}

\author{ {Tuomas Takko} \\
	Department of Computer Science\\
        Aalto University School of Science\\
	00076, Finland \\
	\texttt{tuomas.takko@aalto.fi} \\
	\And
        {Kunal Bhattacharya} \\
	Department of Industrial Engineering and Management\\
        Department of Computer Science\\
        Aalto University School of Science\\
	00076, Finland \\
        \And
        {Kimmo Kaski} \\
	Department of Computer Science\\
        Aalto University School of Science\\
	00076, Finland \\
}

\begin{document}
%\onecolumn
%\firstpage{1}

\title{Modelling exposure between populations using networks of mobility during Covid-19}

\maketitle

\begin{abstract}
The use of mobile phone call detail records and device location data for the calling patterns, movements, and social contacts of individuals, has proven to be valuable for devising models and understanding of their mobility and behaviour patterns. In this study we investigate weighted exposure-networks of human daily activities in the capital region of Finland as a proxy for contacts between postal code areas during the pre-pandemic year 2019 and pandemic years 2020, 2021 and early 2022. We investigate the suitability of gravity and radiation type models for reconstructing the exposure-networks based on geo-spatial and population mobility information. For this we use a mobile phone dataset of aggregated daily visits from a postal code area to cellphone grid locations, and treat it as a bipartite network to create weighted one mode projections using a weighted co-occurrence function. We fit a gravitation model and a radiation model to the averaged weekly and yearly projection networks with geo-spatial and socioeconomic variables of the postal code areas and their populations. We also consider an extended gravity type model comprising of additional postal area information such as distance via public transportation and population density. The results show that the co-occurrence of human activities, or exposure, between postal code areas follows both the gravity and radiation type interactions, once fitted to the empirical network. The effects of the pandemic beginning in 2020 can be observed as a decrease of the overall activity as well as of the exposure of the projected networks. These effects can also be observed in the network structure as changes towards lower clustering and higher assortativity. Evaluating the parameters of the fitted models over time shows on average a shift towards a higher exposure of areas in closer proximity as well as a higher exposure towards areas with larger population.
In general, the results show that the postal code level networks changed to be more proximity weighted after the pandemic began, following the government imposed non-pharmaceutical interventions, with differences based on the geo-spatial and socioeconomic structure of the areas.

%\tiny
% \keyFont{ \section{Keywords:} Collective human mobility, Social Physics, Data-Driven Modelling, Complex Networks, Covid-19} 
\end{abstract}


\section{Introduction}
Studies on human mobility using large scale and longitudinal datasets from real world systems are useful for understanding human behaviour, designing techno-social %of human-engineered 
systems, as well as forecasting different crisis scenarios like pandemics.
%Mobility of individuals, both in terms of short trips and long term migration, can serve as indicators of the state of a society. 
%Short term activity at high level indicates a more active economy or social life whereas low activity can be seen as a sign of economic depression or social segregation~\cite{pappalardo2015using, pappalardo2016analytical}.
Alongside with improved methods for obtaining novel data from mobile phones or other digital devices, human mobility as long range migration and short range trips has been an active field of research. 
The use of mobile phone records data, social media data and other mobile applications has made it possible to capture large and longitudinal datasets of human locations, trajectories and behavioural patterns~\cite{blondel2015survey}. 
This data has motivated the modelling of various aspects of human mobility, such as patterns of their behaviour and trajectories~\cite{gonzalez2008understanding,simini2012universal}, co-locations~\cite{riascos2017emergence}, agent-based systems~\cite{nguyen2011steps,wu2019agent,sakamanee2020methods,jia2012empirical}, and migration~\cite{kang2015generalized,barbosa2018human,yan2017universal,lenormand2016systematic,prieto2018gravity,ren2014predicting}.
Networks are naturally useful for modelling and analysing mobility patterns, as the locations and trips, or entities and social ties, can be considered as nodes and edges, respectively. Networks of telecommunication and proximity have shown characteristic structural properties present in human networks, such as small-worldness and the presence of communities \cite{onnela2007structure}.

Over the years numerous studies have demonstrated the relationship between the mobility and distance, such that the number of trips between two locations has a negative correlation with the distance between them. 
Similarly, when modelling the mobility between two populations, the population sizes have been shown to correlate with the number of trips between the said locations.~\cite{krings2009urban,krings2009gravity,prieto2018gravity}
These gravity type models, inspired by Newton's law of gravity, have also been applied to communications\cite{krings2009urban, krings2009gravity}, social connections and trade \cite{bhattacharya2008international}.
Another commonly used model for describing human mobility is the radiation model, which considers the number of opportunities as the distance variable such that the number of opportunities within the radius of the distance between two locations denominates the probability of an action to another point \cite{kiashemshaki2022mobility}. Such models have been utilized to describe the use of of services, job seeking and migrating.
It should also be noted that there are numerous other external variables affecting the human mobility in addition to the commonly used population sizes and distances, such as available means and cost of transportation or government imposed restrictions \cite{noulas2012tale,souch2021interstates}.

With the emergence of the SARS-CoV-2 pandemic in Finland in early 2020, the authorities imposed non-pharmaceutical interventions (NPIs) in order to contain the spread of the virus and keep the demand for healthcare and hospitalisations on a sustainable level \cite{tiirinki2020covid}.
The global pandemic sparked large interest in studying different models of epidemic spreading as well as strategies regarding the restrictions and the related efficiency \cite{jordan2021optimization}.
These NPIs included such methods as social distancing  \cite{soucy2020estimating, thurner2020network}, contact tracing \cite{barrat2021effect} as well as restrictions like school closings and remote work guidelines \cite{jordan2021optimization,haug2020ranking}, which affect the overall daily behaviour of people, should they comply with them.
In addition to the restrictions imposed by the government, the self-imposed social distancing of health-aware people had an effect on their mobility as well as on the spread of the virus.
As stated in the study by Chang and coauthors~\cite{chang2021mobility}, the models for the spread of the epidemic will need to take into account the effects of changes in mobility.
These changes, caused by restrictions or general awareness, have been studied at the level of countries \cite{bergman2020correlations,askitas2020lockdown, sulyok2020community, sadowski2021big} and at a more microscopic level such as cities and grids  \cite{badr2020association, bergman2020mobility, kim2021impact, chang2021mobility,levin2021insights}.
Worth noticing is that the dynamics of epidemic spreading have been shown to be more complex than just the contacts caused by activity \cite{badr2021limitations} and the estimated epidemiological effects of different NPIs have been shown to be model %reliant 
dependent \cite{chin2021effect}.

%Overall introduction to the paper
In this study we investigate the human behaviour before and during the Covid-19 pandemic in Finland using the daily activities as a proxy for estimating the possible contacts between the populations living in different postal code areas. The overarching objective of this study is to use modelling to show large scale changes in exposure induced by human mobility during the pandemic using postal area level networks. Our first research objective is to use the aggregated daily activity data for creating models of possible exposure-networks (or contact networks).
The second objective is to quantify the changes in this system during the first two years of the pandemic in comparison to the preceding prepandemic year 2019.
Finally, we aim to show a relationship between the government imposed NPIs and the exposure networks, using data from \cite{hale2021global}.

While many other studies have used the activity-type aggregated data from companies such as SafeGraph, CueBiq, Google or Apple (see \cite{askitas2020lockdown, bergman2020correlations, sulyok2020community, bergman2020mobility}) for various countries and regions, we seek to obtain new insight to the relationship of activity and the NPIs by using postal code and cell grid level data on a daily time resolution. 
This contains more micro level activity than the community level reports used in some studies, but less information than the individual mobility traces used in some other studies.
The data we use is obtained from Telia, a telecom company operating in Finland, and it contains aggregated daily activity data of mobile phone users.
Similar data has been used in other studies such as \cite{wetter2020private, willberg2021escaping, kiashemshaki2022mobility,levin2021insights}, but not in the same type of modeling paradigm.
% OTHER STUDIES AND DIFFERENCES
The methodology of using gravity and radiation type models for predicting the movement of people between two areas with varying population, number and diversity of services \cite{chong2020economic}, wealth or similarity \cite{yabe2022behavioral} has been investigated with different datasets of CDR or GPS locations in the form of origin-destination networks.% in multiple countries and geographical areas. 
%Kiashemshaki et al.\cite{kiashemshaki2021mobility} also used a dataset from the same mobile service provider as in our case in the context of trips between cities in Finland as an origin-destination matrix.
Our study considers a bidirectional relationship between postal code areas, where the populations in these areas do not require visiting the other's geographical area to form an edge.
Thus, the bidirectional edges represent the interaction of exposure in a broader setting than directed trips between an origin area and a destination area. 
%The interpretation of an edge in this study also differs from the previously mentioned studies, but shares a similarity to studies with communication as the underlying phenomenon\cite{krings2009gravity}.

The rest of the manuscript is divided into four sections, Methods, Results, Discussion, and Conclusions. First, we present the daily activity data along with the pre-processing steps done to make the data into sets of networks for periods of time. From these networks we construct two types of models, a gravity-type model and a radiation-type model, by using postal code area level data. In the Results section we present the findings on the fitting of the two classical models and the changes in the resulting networks over the course of years 2019 to 2021. In the Discussion section consider the implications of our findings and discuss potential future research. 
%Finally, we discuss the findings and the future research objectives.
In the Conclusions section we present overall concluding remarks.

\section{Material and Methods}
In this section, we describe the empirical dataset and explain the construction of the weighted bipartite projections, the measures used to investigate the temporal changes in the network and finally the proposed model for predicting the likelihood of social connections between areas using geospatial and socioeconomic information.

\subsection{Data}
The data used in this study is obtained from Telia, a telecommunications company operating in Finland, Sweden and Denmark.
This data, called "Crowd Insights", contains aggregated human activities for each day in the format of home and grid location matrix.
Each mobile phone user's home location is marked with its postal code area, where the user stayed longest before 9 am, i.e. the place the user woke up in the morning.
The activity of each user is recorded with the binary (0 or 1) if the user spent at least 20 minutes during the day in that particular grid location.
This data is then anonymised and aggregated for each postal code area, thus generating a matrix where during one day a postal code area has an amount of visits or mobile activities to each grid cell. The grid was defined by the telecom areas, ranging in size from 500 metres by 500 metres to 2 kilometres by 2 kilometres depending on the density of users in the area.
In order to protect the privacy of less populated areas, activities with less than 5 people are redacted from the dataset.
The dataset contains the activities during the periods 1.1.2019--23.9.2019 and 1.2.2020--31.3.2022, with some days being removed due to low signal quality.
For reducing the possible inaccuracy caused by the incomplete yearly sets and the seasonal trends therein, we create uniformly sized subsets where each date is contained in the set of each year.
By filtering these subsets, we obtain 219 dates for each year from 2019 to 2021.
These subsets are referenced as yearly sets in this manuscript. 

In this study we consider a subset of the data by choosing only the activities that 
are both stemming from and occurring within the capital area of Finland, meaning the municipalities of Helsinki, Vantaa, Espoo and Kauniainen.
In other words, an accepted activity is performed by a person waking up in the chosen geographical area and performing the activity in the grids of the area.
Thus, we consider the capital area as a separate subsystem from the rest of the country, even though the data contains people moving in and out of the system.
Postal code areas in the chosen geographical area are highly populated, but also smaller in size than the average postal code areas in the country.
%The choice of geographically smaller area also contains a larger quantity of short term activity when compared to activities taking place between areas hundreds of kilometres apart. 

Filtering the data results in 167 postal code areas as home areas and 1444 grids as the areas for performing mobile activities.
This data can be presented as a bipartite network, where for each day we consider the home areas and grid areas as separate classes of nodes and the links are the number of people who have performed the mobile activity.
In the scope of this paper, we are investigating the difference in human behavior during the years 2019 to 2022 by constructing networks of the postal areas and analyzing it at the graph level using various time windows for the activity aggregation.
The aim is to use the commonalities between activities of various areas as a proxy for contact as well as functional activities such as work and thus form a network of connected postal areas.

In addition to the activity data, we use socioeconomic data on the same postal code area level as variables in the model.
The data is obtained from Statistics Finland \cite{paavoreference} and is openly available.
In this study we are not considering the changes in the socioeconomic values for each postal code area as the changes can be considered to be minor during the timespan of the study and the reporting of postal code areas changes between the years in the Statistics Finland data.
As the final source of information we use the dataset by Hale et al.~\cite{hale2021global} for obtaining a numerical index on the level of mobility and other activity related restrictions in Finland during the pandemic.

For the analysis of the daily contact networks between the postal code areas we calculate a set of graph theoretical properties with the intention of observing changes and quantify some of the characteristics of the networks.
Daily population activities follow a weekly schedule as well as public holidays and holiday periods, resulting in relatively high variance throughout the year. 
We address this by analyzing the network properties on daily networks as moving averages as well as networks aggregated as averages of 7-day periods or as years.
The dataset contains empty gaps, namely the time between 24.09.2019 and 31.01.2020, which we address by comparing only the common dates during each year when doing comparisons between the years. Otherwise these empty periods of time are visualized as discontinuities.

For measuring the changes in the network and reinforcing the reasoning behind the models, we calculate the average clustering coefficient, assortativity of degree correlations and weighted degree distributions. Then we compare the resulting weighted networks to the distances and population sizes used in Eq. \ref{eq:gravity} and Eq. \ref{eq:radiation} without fitting the data beforehand.
For the calculation of the average clustering coefficient and the degree assortativity we utilize the NetworkX library\cite{hagberg2008exploring} in Python 3.7.

\subsection{Models}
In this section we describe first the process of constructing the networks of exposure from human mobility using the above described empirical data and then fitting two types of models to simulate the interactions between postal code areas, i.e. the nodes of the network. 
%These bipartite networks of activity between cellphone grids and postal code areas are projected into daily exposure networks, acting as proxies for contacts between postal code areas. 
%We use these networks as sets of egocentric networks
We use the edges of these projected networks for fitting a gravity type model and a radiation type model using postal code level population data. The changes in the structure of these networks over time are analysed using standard graph theoretical measures.

\subsubsection{Bipartite network projection}

\begin{figure}[h!]
    \centering
    \includegraphics[width=1\textwidth]{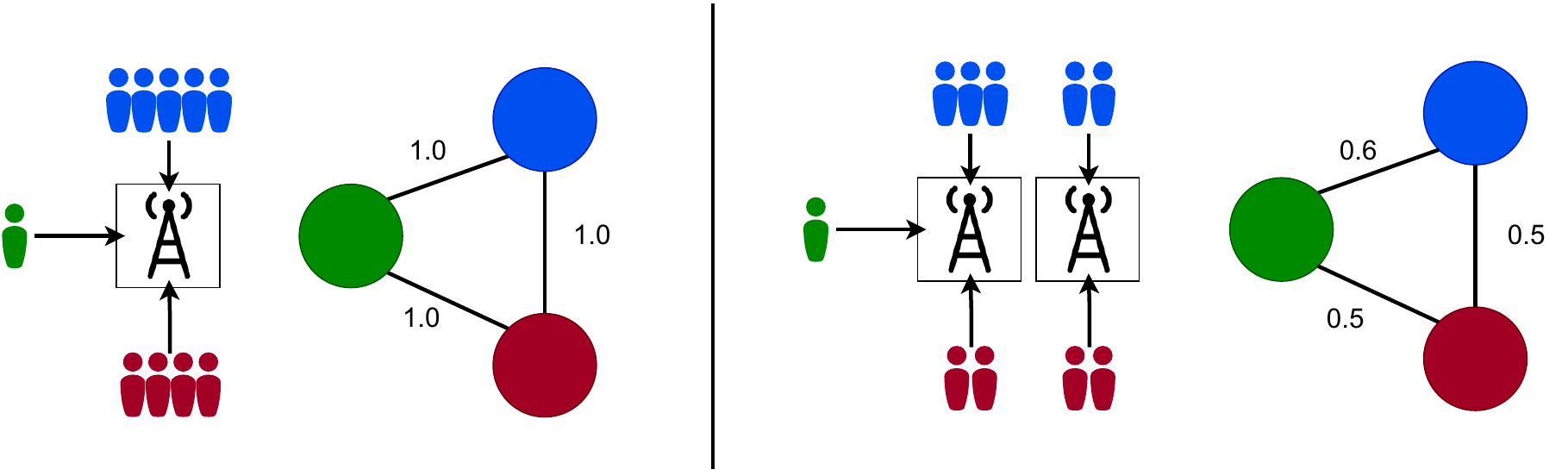}
    \caption{Two examples of the weighted exposure network construction for three example postal code areas, red, blue and green. The first example consists of only one grid area and the second has 2 grid areas. Both examples assume no other activities. Each person represents one activity from the respective postal code area to the grid area connected by an arrow. The 3 nodes represent the corresponding projection with edge weights illustrated next to the edges and calculated using Eq. \ref{eq:weight}.
    }\label{fig:figure1}
\end{figure}

The temporal activity data forms a bipartite network in the form of an adjacency matrix $A$ of size $167 \times 1444$, where each postal area (row) has a number of activities in each geolocated grid (column).
From this matrix we calculate a weighted one mode projection by defining the edge weight function $W_{ij}$ for the postal code areas $i$ and $j$ as the sum of products over each grid cell:
%From this matrix we calculate a weighted one mode projection (WOMP) on the postal areas using a weighting function \ref{eq:weight}. For two postal code areas $i$ and $j$ the WOMP edge weight $W_{ij}$ would be calculated as the sum of products over each grid cell

\begin{equation}\label{eq:weight}
   W_{ij}(t) = \sum_g \frac{w^g_i(t)}{\sum_{z}w^z_i(t)}\frac{w^g_j(t)}{\sum_{z}w^z_j(t)}, 
\end{equation}
where $w^g_a(t)$ is the number of activities, i.e. people visiting the grid cell $g$ from the postal code area $a$ at time $t$ and the denominator is the sum of all activities from the corresponding postal code area.
The multiplication ensures that both postal code areas, $i$ and $j$, have activity in the particular grid cell during that day. The weight $W_{ij}$ is bounded between 0 and 1, where 0 would indicate no co-occurring activities and value 1 would indicate full co-occurrence to a single grid cell, e.g as depicted in Fig. \ref{fig:figure1}.
A small value indicates a low probability for picking a random activity from each postal code area such that both of the activities were performed in the same grid cell.
This way we obtain a weighted projection between the postal code areas that can be used as a proxy for contacts between the postal code areas.
An example of the dynamics of the function \ref{eq:weight} is depicted in Fig. \ref{fig:figure1}.
Using this method for the duration of the empirical data we obtain $S=1031$ daily networks, in which the nodes are the only permanent structure.

\subsubsection{Modelling networks of exposure}

The main research goal of this paper is to create a characteristic model for predicting and understanding the resulting bipartite network projections (Eq.~\ref{eq:weight}) that represent the contacts between the postal code areas due to the co-occurrence of individual mobile activities or mobilities. For this we use two classical models: a gravity-type model and a radiation-type model \cite{krings2009gravity,kiashemshaki2022mobility,jia2012empirical,nguyen2011steps}. 
In case of the gravity-type model, we consider the exposure between populations in two postal code areas, $f_g(i,j) = W_{ij}$, to be  to dependent increasingly on the size of the two populations and decayingly on the distance between them, as follows 
\begin{equation}\label{eq:gravity}
f_g(i,j)=  C\frac{(p_{i}p_{j})^{B_{p}} }{ d_{i,j}^{B_{d}} },
\end{equation}
where $C$ is a constant, $B$ is the parameter specific exponent, $p_i$ and $p_j$ are the population variables of the postal code areas $i$ and $j$, and $d_{ij}$ is the "crow-flight" distance between the centroids of the land areas (excluding waters) of these postal areas, measured in meters. %, and $p_i$ and $p_j$ is the population variable of the postal code area $i$ and $j$.
%For distance, we calculate the crow-flight distance between the centroids of land areas (i.e. excluding waters) of each postal area in meters. 
This is an assumption since the population is not evenly distributed and the real world distances can be longer due to the transportation infrastructure. 
As in literature we consider our gravity-type model symmetrical, i.e. the exposure or the individual mobility is the same from $i$ to $j$ as from $j$ to $i$. 
%The population based variables are obtained from a publicly available dataset of Statistics Finland \cite{paavoreference} for the year 2019 and we make an assumption that they do not change significantly over the investigated time period. 
%The variables we use from the dataset~\cite{paavoreference} for the year 2019 are total population, number of workplaces, number of services, and number of households.

The second model we evaluate with the empirical networks (Eq.~\ref{eq:weight})) is based on the radiation-type models, which in general are formulated as directed interactions. However, for the problem of this study we consider the interactions symmetrical as in the gravity-type model introduced above. 
The model is based on the notion of number of possible opportunities within the radius equal to the distance between two areas centered on the source area.
Here the radiation model for exposure, $f_r(i,j)=W_{ij}$, is constructed using the following modified form of the standard model:
\begin{equation}\label{eq:radiation}
f_r(i,j)= \frac{p_ip_j^{B_{p}}}{[(p_i+r_{d_{i,j}}+p_j)(p_i+r_{d_{j,i}}+p_j)]^{B_{r}}},
\end{equation}
where $r_{d_{i,j}}$ is the sum of population variable within the radius $d_{i,j}$ from the area $i$ excluding the populations in areas $i$ and $j$, and $B$ is the parameter used for fitting the model to the data.
The population variable under radius, $r_{d_{i,j}}$, was obtained by calculating the fraction of the area under the radius $d_{i,j}$ for each postal code area and multiplying each population variable by the fraction of area covered. Here it is assumed that the population, services or workplaces are evenly distributed.
We use this approximation due to lack of more detailed data.
Prior to fitting the models, we filter out the postal code areas with less than 500 inhabitants, which from our empirical dataset removes only seven areas that are mostly uninhabited industrial zones.

Both the gravity and radiation type models are fitted to the weights of the edges of the empirical 7-day average networks and the yearly average networks with a linear regression utilizing ordinary least squares (OLS) for obtaining the parameter values. To use a linear regression model we change the models in logarithmic form such that the gravity type model is of the form

\begin{equation}
   log(W_{ij}(t)) =  B_p*log(p_ip_j)-B_d*log(d_{i,j})
   \label{eq:regressionfuncgravi}
\end{equation}
while the radiation type model is of the form
\begin{equation}
   log(W_{ij}(t)) =  B_p*log(p_ip_j)-B_r*log((p_i+r_{d_{i,j}}+p_j)(p_i+r_{d_{j,i}}+p_j)).
   \label{eq:regressionfuncradi}
\end{equation}
The statistics of the fitted exponents ($B_{p}$, $B_{p}$,$B_d$,$B_{r}$) are shown in Table \ref{tab:table1} and in Fig.~\ref{fig:figure5} the values are visualized over the time period of our dataset. 

After fitting and evaluating the gravity and radiation models, we investigate whether combining additional geospatial and population level variables to the gravity type model would give a better estimate of the edges of the network. These additional variables were chosen by evaluating the edges with largest error.
The chosen variables are the population density, obtained from the socioeconomic Statistics Finland dataset~\cite{paavoreference}, and a simple distance via public transportation (i.e. train and metro). The distance via public transportation was obtained by using the geographical coordinates of each train and metro station and their interconnections within the four municipalities. For each pair of postal code areas we calculate the distance from the centre of the area to the nearest station and then calculate the shortest path length via public transportation between the corresponding nearest stations. In the case both areas have the same nearest station, the distance is set as the distance between the two area centres. The final form of this extended model $f_e(i,j)$ is as follows

\begin{equation}\label{eq:gravityextend}
f_e(i,j)=  C\frac{(p_{i}p_{j})^{B_{p}}*(\hat{p}_{i}\hat{p}_{j})^{B_{\hat{p}}}}{ d_{i,j}^{B_{d}}*  \hat{d}_{i,j}^{B_{\hat{d}}} },
\end{equation}
where $\hat{p}$ is the population density and $\hat{d}$ is the distance via public transport.
The extended model was fitted similarly to Eq. \ref{eq:regressionfuncgravi} and Eq. \ref{eq:regressionfuncradi}.

\section{Results}
In this section, we will construct the projected empirical networks, analyze them in terms of general attributes and their time-dependent structural changes as well as fit the above introduced three models to the data.
In total, the dataset contains 1031 days of individual mobile activities in the original grid cells which we project to daily postal area level weighted exposure networks using the formula in Eq. \ref{eq:weight}. The aggregated networks were constructed such that each edge weight was the mean of the corresponding edge within the chosen time frame.
The weekly averaged networks were constructed as 7-day consecutive bins and the yearly averaged networks are constructed using overlapping dates during the years 2019, 2020 and 2021 in the data (dates from 01.02. and 23.09.).
With this binning we obtained 146 weekly networks and the yearly overlapping dates contain 219 days, which were used in the aggregated yearly networks.

\begin{figure}[h!]
    \centering
    \includegraphics[width=1\textwidth]{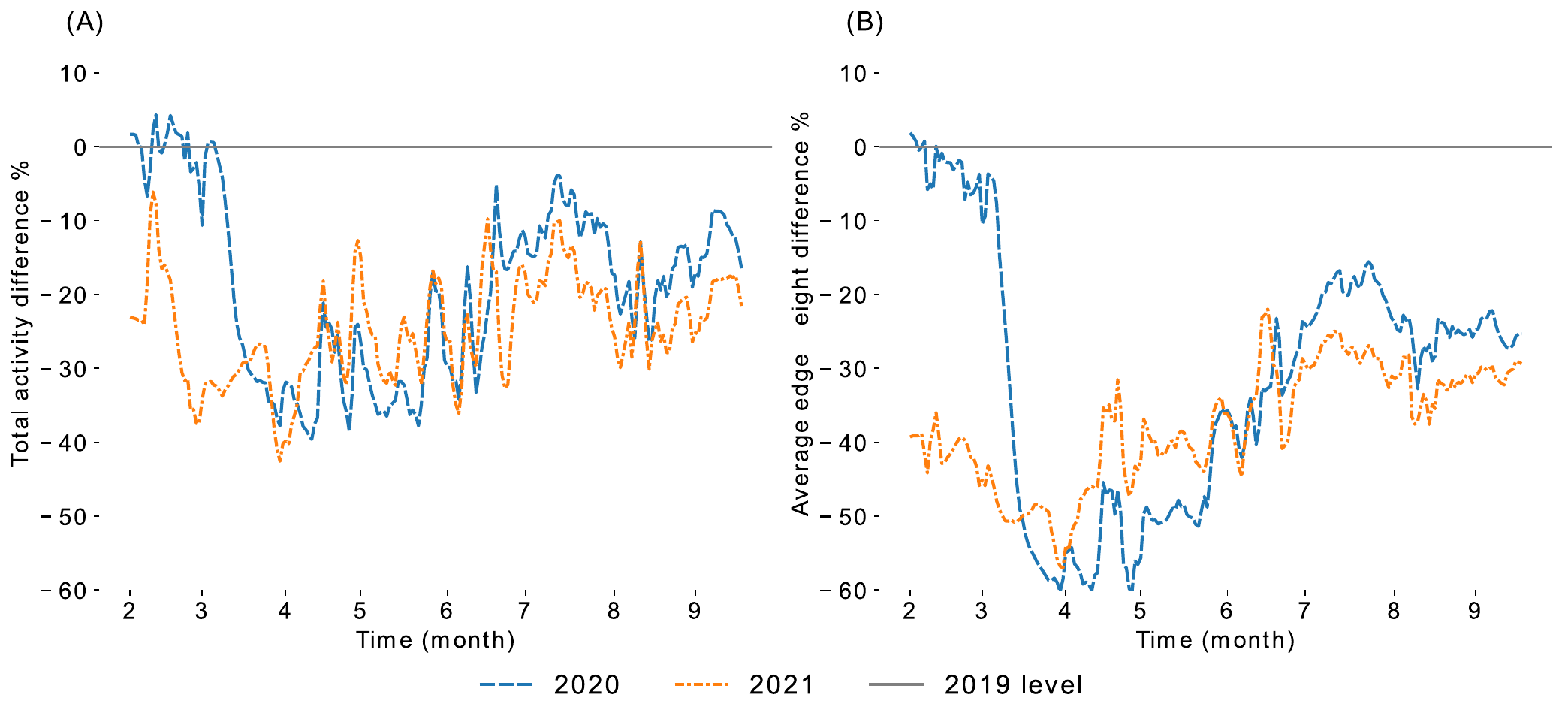}
    \caption{
    Proportional percentage difference of the total activity in the empirical bipartite networks (A) and of the average edge weight in projected exposure-networks (B) between the overlapping dates for the years 2020 and 2021 to the pre-pandemic year 2019 as 7-day moving averages.
    }\label{fig:figure2}
\end{figure}

\begin{figure}[h!]
    \centering
    \includegraphics[width=1\textwidth]{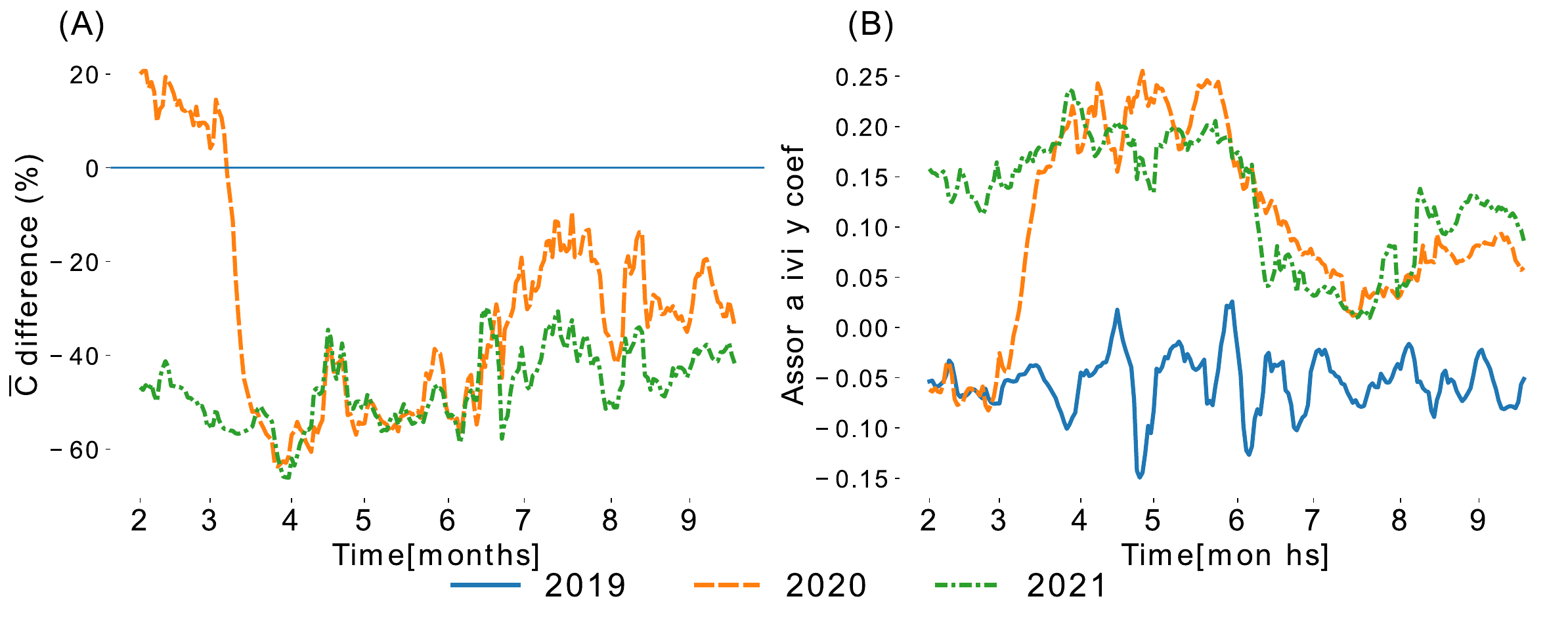}
    
    \caption{7-day moving averages of measures of the projected networks during the yearly overlapping dates in the dataset: (A) Average weighted clustering coefficient as proportional difference to the same time in the year 2019, (B) weighted degree assortativity correlation coefficient. The networks in both (A) and (B) are filtered by edge weight to have the 90-percentile of edges based on edge weight.
    }\label{fig:figure3}
\end{figure}

\begin{figure}[h!]
    \centering
    \includegraphics[width=1\textwidth, trim={2cm 1cm 2cm 1cm}, clip]{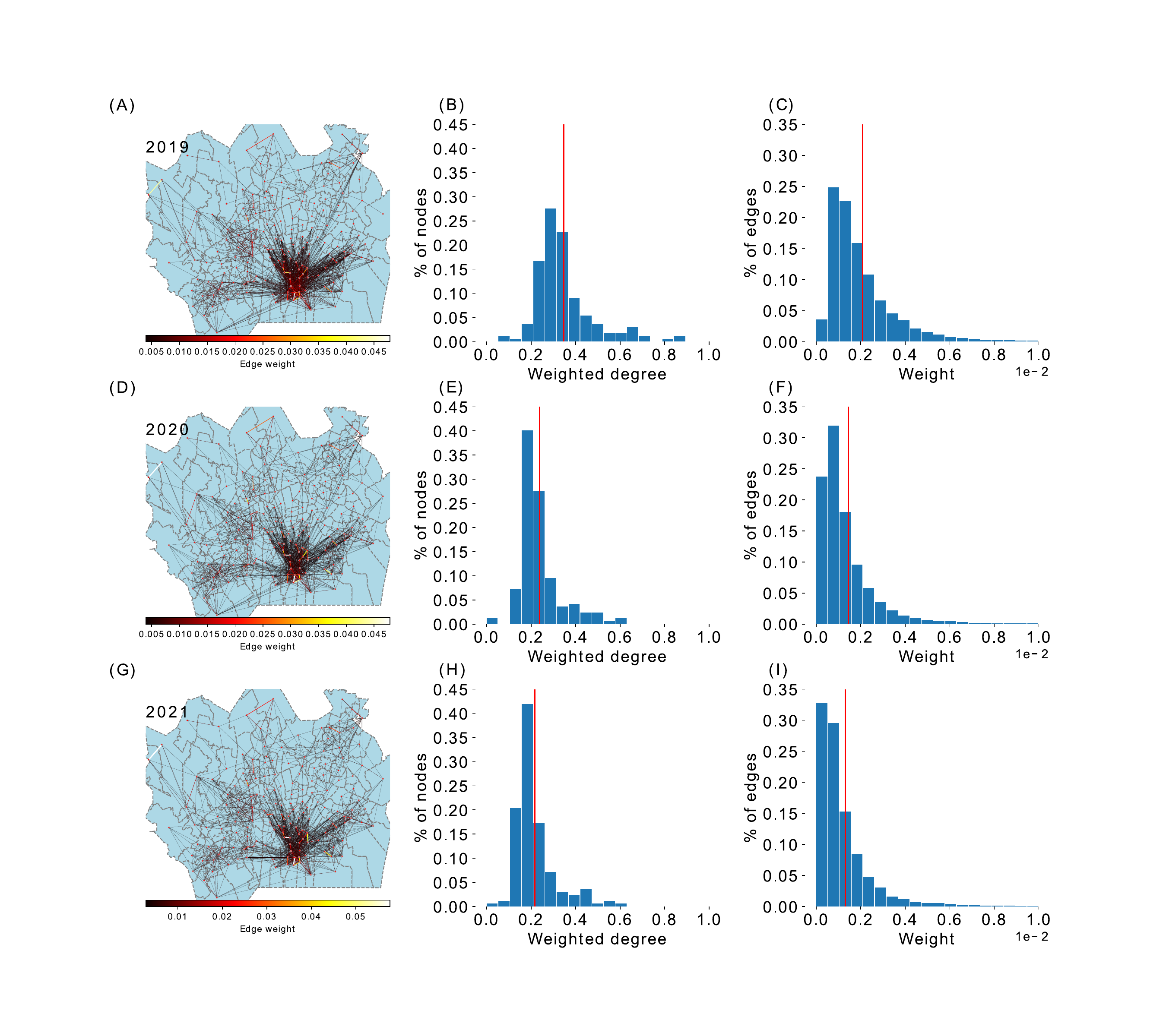}
    
    \caption{
    The yearly projected networks with the mean edge weights for the years 2019 (A-C), 2020 (D-F) and 2021 (G-I).
    In sub-figures A, D and G the network is shown as 90th percentile of the edges filtered by edge weight. The width and colour depict the weight of the edge and the node size depicts the weighted degree. The histograms in sub-figures B, E and H depict the weighted degree distribution of each yearly network. The histogram has 20 bins and the red vertical line depicts the mean. The rightmost histogram depicts the distribution of edge weights.
    }\label{fig:figure4}

\end{figure}

%first figure
In Fig. \ref{fig:figure2} the total daily activity and the average edge weight of the pandemic years 2020 and 2021 are visualized in comparison to the same dates during pre-pandemic year 2019 as a moving average. Both these measures show similarities in terms of the overall difference to the pre-pandemic year.  Both the seasonality and public holidays can be seen as peaks in the overall activity, whereas the same peaks are not as prominent in the average edge weight. Also worth noticing is the proportional difference, as the total activity at points exceeds the 2019 level, the average edge weight does not.
Comparing the time series of the activity measure and the average edge weight to the average containment index from the NPI dataset~\cite{hale2021global} shows a negative correlation as would be expected if the population is compliant or epidemic aware. Interestingly the correlation for the average edge weight is stronger than for the total activity, as indicated by the correlation coefficients $\approx -0.76$ and $\approx -0.56$, respectively. The correlation coefficient between the same measures is $\approx 0.72$. All correlation coefficients are shown in Table~\ref{tab:table2}.
%next figure
The resulting aggregated yearly networks and the related weighted degree and edge weight distributions are shown in Fig.~\ref{fig:figure4}.
As can be expected from the proportional comparisons in Fig.~\ref{fig:figure2}, the means of the distributions shift towards 0 during the pandemic. The highest weighted degree becomes also lower, i.e. $\approx 0.873$ to $\approx 0.622$ and $\approx 0.603$).
For clarity of visualization, the yearly networks are depicted using the four strongest edges for each node with the node width and colour representing the weight in Fig.~\ref{fig:figure4}. 
A visually noticeable difference is the decrease in the number of longer distance edges to the geographical center (central Helsinki) taking place after 2019. 
Utilizing the Clauset-Newman-Moore greedy modularity maximization~\cite{clauset2004finding} for clustering the postal code areas to communities in the aggregated yearly networks using the 95-percentile of the edges based on weight shows that the number of detected communities declines from 16 during pre-pandemic 2019 to 4 and 5 during pandemic 2020 and 2021.

Next we measure the average clustering coefficient and degree assortativity in the filtered 90-percentile graphs and perform a preliminary analysis for the models by calculating the Pearson correlation between the edge weights of the full graph and the corresponding population level variables as described in Eq.~\ref{eq:gravity} and Eq.~\ref{eq:radiation}. The time series are shown in Fig.~\ref{fig:figure3}.
The proportional difference in the weighted average clustering coefficient in Fig.~\ref{fig:figure3}A shows similar change as the average edge weight. 
The Pearson weighted degree assortativity, depicted in Fig.~\ref{fig:figure3}B, increases during the pandemic, which indicates that the edges in the filtered aggregate networks during the pandemic are more often found between postal code areas with similar weighted degree. This effect can also be seen in the network visualizations in Fig.~\ref{fig:figure4}, where the strongest edges are no longer between the central area and areas further from it. Also, this correlates with the notion of larger communities detected during the pandemic. The result remains uniform when replacing the edge weight with $1$ in the filtered network.

\subsection{Evaluating the models} 
\begin{figure}[h!]
    \centering
    \includegraphics[width=1\textwidth]{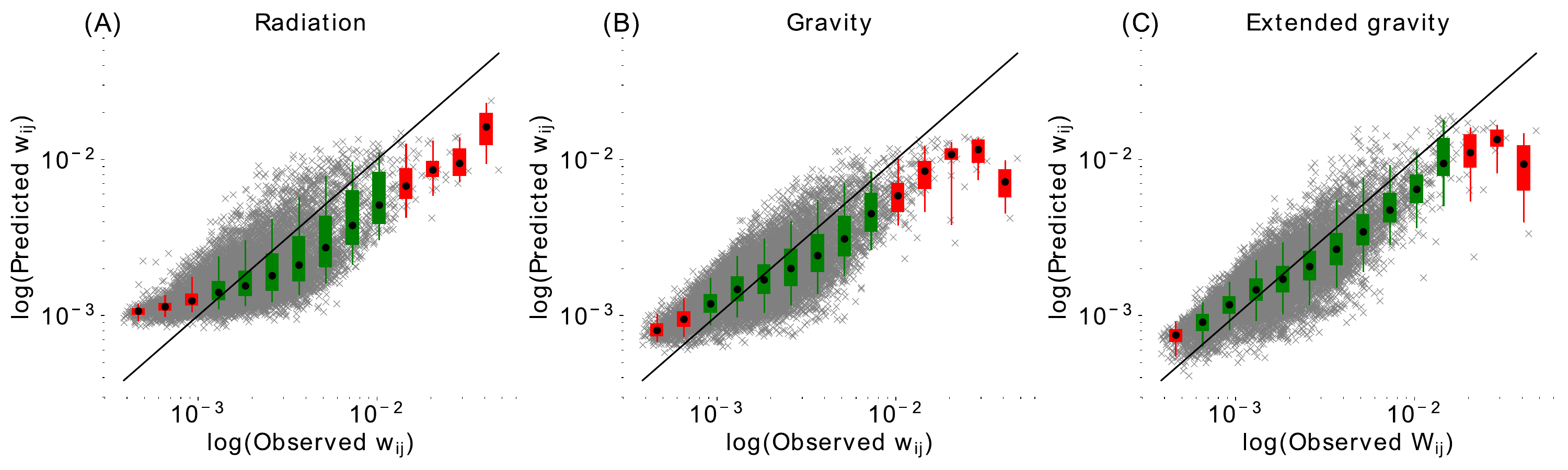}
    \includegraphics[width=1\textwidth]{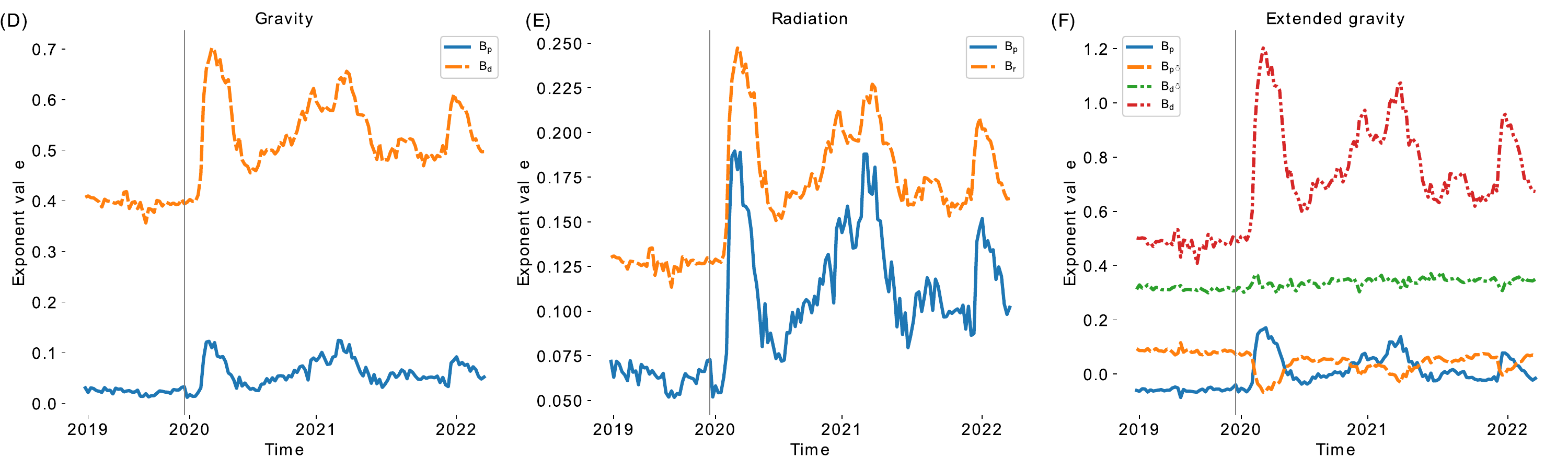}
    \caption{Fitting of the aggregated networks to the 3 models. (A-C) The predicted model values plotted against the observed empirical values for the network of 2019. Each point denotes an edge in the network and the straight line indicates the perfect fit. Points closer to the line have smaller error and points above the line are overestimates and vice versa. The points are visualized also in equally sized bins, depicting the mean (dot), the 25th percentile (box) and the 95th percentile (line connected to box). The colour of the box is green if the 95th percentile contains the perfect fit to the observed data (straight line).
    The rest of models fitted to aggregated yearly networks are shown Fig.~\ref{fig:figureS1} in Supplementary Material. 
    (D-F) The fitted model exponents for the whole duration of the data with 7-day aggregated networks. The vertical line denotes the %empty period in the 
    missing data between 24th September 2019 and 31st January 2020.
    }\label{fig:figure5}
\end{figure}

For the performance evaluation of our three models we first fit them to the aggregated yearly networks with the empirical data. Then we perform the evaluation by calculating the errors using symmetric mean absolute percentage errors (SMAPE) and root mean squared error (RMSE):
\begin{equation}
    RMSE = \sqrt{\frac{\sum^n (W_f - W_o)^2}{n}}, SMAPE = \frac{1}{n}\sum^n \frac{\left|W_f - W_o\right|}{\left|W_f\right|+\left|W_o\right|},
\end{equation}
where $n$ is the number of edges, $W_f$ is the predicted edge weight and $W_o$ is the observed edge weight. The results for the fitted exponents and errors are shown in Table~\ref{tab:table1}. The predicted values obtained for the radiation, gravity, and extended gravity models are plotted against the observed values for the year 2019, depicted in Figures~\ref{fig:figure5} (A), (B), and (C), respectively.
As anticipated by the correlations to the unfitted models, the gravity model performs better than the radiation model in predicting the edge weights and the extended gravity model performs the best. Same can be seen in the models' $R^2$ values, which indicate the proportion of variance explained by the variables. All the three models have $R^2$ values over $\geq 0.5$ in 2019 with interestingly an increase during the pandemic years. The difference in proportional error (SMAPE) between the three models remains consistent in all three aggregated yearly networks. 
The values of RMSE show that in the pre-pandemic year 2019 the radiation model shows the lowest error, but in the subsequent years the error shows an increasing trend for all three models. The error metrics over time for the weekly networks are shown in Fig.S2%~\ref{supp-fig:figure6}.

The fits of the models in Fig.~\ref{fig:figure5} illustrate the error. All three models overestimate the weakest edges and underestimate the strongest edges with varying magnitude, the radiation model having the largest error.
%Model coefficient interpretation
While the absolute values between fitted exponents of the same model are not comparable due to the different units in the variables, the proportional differences show changes in the aggregated networks.
The exponents of gravity model indicates a larger significance on the exponent for population $B_p$ after 2019 as it increases in proportion to the distance exponent $B_d$. 
The distance exponent shows the previously discussed weakening of longer distance edges by increasing with each year.
The fitted exponents of the radiation model show a similar change with the increasing weight of $B_p$.
The exponent for population within radius, $B_r$, increases but not in proportion to $B_p$.
When including the population density in the extended gravity model, the fitted population size exponent becomes negative in 2019 and increases during the pandemic while $B_{\hat{p}}$ decreases. $B_d$ increases similar to the basic gravity model, decreasing the weight of long distance edges. The exponent for the distance via public transportation, $B_{\hat{d}}$ remains in the same order of similar magnitude throughout the three years, i.e. from prepandemic 2019 till pandemic 2020 and 2021.

%Model exponents over time
After fitting and analyzing the yearly aggregated networks, we investigate the full duration of our dataset with the weekly aggregated networks.
The fitted exponents over the the course of the dataset are shown in Fig.~\ref{fig:figure5}.
The gravity model over time shows the two exponents $B_d$ and $B_p$ behaving similarly, while their relative difference changes. Similar shape can be seen in the case of radiation model but with more extreme relative changes. In the case of the extended gravity model the two additional variables show a negative relationship to exponents of the basic gravity model. The point in time when the exponent $B_p$ becomes positive can be seen to occur at the beginning of the pandemic.
Comparing the fitted exponents to the average edge weight over time shows high correlations as can be expected. 
Consequently, most of the exponents share a high correlation to the NPI restriction index. The outliers in terms of the correlation strength is the added variable $\hat{d}$ in the extended gravity model.
These correlations are listed in Table~\ref{tab:table2}. 
In order to investigate the effects of particular types of NPIs, we fit the 8 sub-indices related to the restriction stringency in the OxCGRT dataset~\cite{hale2021global} to each of the fitted exponents ($B_p$, $B_d$, $B_{\hat{p}}$ and $B_{\hat{d}}$) of our extended gravity model (~\ref{eq:gravityextend}) using a similar OLS regression as before. The predicted values of each of these exponents are shown in Fig.~\ref{fig:figureS3} and the related coefficients in Table~\ref{tab:table3}. %~\ref{supp-tab:table3}.
The results show varying coefficients and p-values for each different NPI sub-index depending on the exponent they were fitted to. Across the four models fitted extended model's exponents, the stringency subindex ´´Restrictions on gatherings'' has a constantly low p-value and standard error. In case of school and workplace closures, restrictions on internal movement and cancellation of public events also show significant p-values ($<0.05$). In addition, the closures on public transport have an effect on $B_{\hat{d}}$, as expected.

Finally, comparing the predicted networks to the corresponding observed networks reveals the lacking aspect of gravity and radiation type models in the context of networks. The previously discussed errors and model fits indicate that the models capture the aspect of exposure between postal code areas. This can also be seen when comparing the average edge weights of the predicted networks and observed ones, as the differences between the two values are of the order of $10^{-14}$. However, a difference arises when comparing the average weighted clustering coefficient and the average weighted degree assortativity coefficient. The model produces networks with higher assortativity and clustering than observed in the empirical networks. 

\def\arraystretch{1.5}
\begin{table}[h]
    \centering
\begin{tabular}{l|c|c|c|c }
        \multicolumn{2}{c}{}  & \textbf{2019} & \textbf{2020} & \textbf{2021} \\ \hline
        \multirow{6}{*}{\textbf{Gravity}}& $C$ & 0.4949 ($\pm$8.3$*10^{-2}$) & 1.7864 ($\pm$9.2$*10^{-2}$) & 1.9716 ($\pm$9.8$*10^{-2}$) \\
        & $B_p$ & 0.0212 ($\pm$4.0$*10^{-3}$) & 0.0316 ($\pm$4.0$*10^{-3}$) & 0.0575 ($\pm$4.0$*10^{-3}$) \\
        & $B_d$ & 0.3913 ($\pm$2.0$*10^{-3}$) & 0.4980 ($\pm$3.0$*10^{-3}$) & 0.5412 ($\pm$3.0$*10^{-3}$) \\
        & SMAPE & 0.1618 & 0.1788  & 0.1891 \\
        & RMSE & 1.331$*10^{-3}$  & 1.337$*10^{-3}$  & 1.398$*10^{-3}$  \\
        & $R^2$ & 0.6  & 0.665  & 0.674 \\
    \hline
         \multirow{6}{*}{\textbf{Radiation}} & $C$ & -0.9098 ($\pm$8.8$*10^{-2}$) & 0.3904 ($\pm$9.5$*10^{-2}$) & 0.4898 ($\pm$0.102) \\
         & $B_p$ & 0.0608 ($\pm$4.0$*10^{-3}$) & 0.0816 ($\pm$4.0$*10^{-3}$) & 0.1117 ($\pm$4.0$*10^{-3}$) \\
        & $B_r$ & 0.1256 ($\pm$1.0$*10^{-3}$) & 0.1672 ($\pm$1.0$*10^{-3}$) & 0.1824 ($\pm$1.0$*10^{-3}$) \\
       	& SMAPE & 0.1810  & 0.1937  & 0.2055  \\
       	& RMSE & 1.145$*10^{-3}$  & 1.358$*10^{-3}$  & 1.401$*10^{-3}$  \\
       	& $R^2$ & 0.5  & 0.605  & 0.618  \\
     \hline
          \multirow{8}{*}{\textbf{E.Gravity}}& $C$ & 0.8885 ($\pm$0.083) & 2.513 ($\pm$0.0965) & 2.7746 ($\pm$0.103) \\
          & $B_p$ & -0.0605 ($\pm$4.0$*10^{-3}$) & -0.0092 ($\pm$5.0$*10^{-3}$) & 0.0190 ($\pm$5.0$*10^{-3}$) \\
         & $B_{\hat{p}}$& 0.0855 ($\pm$3.0$*10^{-3}$) & 0.0382 ($\pm$3.0$*10^{-3}$) & 0.0352 ($\pm$3.0$*10^{-3}$) \\
        & $B_d$ & 0.4774 ($\pm$9.0$*10^{-3}$) & 0.7085 ($\pm$0.01) & 0.7815 ($\pm$0.011) \\
        & $B_{\hat{d}}$ & 0.4774 ($\pm$9.0$*10^{-3}$) & 0.3331 ($\pm$0.011) & 0.3532 ($\pm$0.012) \\
       	& SMAPE & 0.1465  & 0.1684  & 0.1804 \\
       	& RMSE & 1.240$*10^{-3}$  & 1.297$*10^{-3}$  & 1.364$*10^{-3}$  \\
       	& $R^2$ & 0.668  & 0.697  & 0.703 \\
    \end{tabular}
    \caption{
    The fitted coefficients for the yearly average networks. 
    The means are obtained by fitting the models for the full yearly averaged networks with postal code areas of population less than 500 being removed from the network. 
    The number in brackets is the standard error for the coefficient. SMAPE and RMSE measure the error of the fitted model and lower number indicates a better fit. 
    }
    \label{tab:table1}
\end{table}

\def\arraystretch{1.5}
\begin{table}[h]
    \centering
\begin{tabular}{l|c|c|c }

         &\textbf{Variable} & \textbf{$\overline{W}$} &  \textbf{$\overline{NPI_c}$} \\ \hline
         \multirow{2}{*}{\textbf{Data}}& $\overline{W}$  &  & -0.8235  \\
         & $\sum_i \sum_g w_{i}^{g}$ & 0.7146 & -0.6682 \\
\hline
      \multirow{2}{*}{\textbf{Gravity}}  & $B_p$ & -0.9206 & 0.6997 \\ 
        & $B_d$ & -0.9859 & 0.8299  \\
    \hline
        \multirow{2}{*}{\textbf{Radiation}}&$B_p$ & -0.9474 & 0.7358\\
        &$B_r$ & -0.9851 & 0.8436\\
     \hline
        \multirow{4}{*}{\textbf{E.Gravity}} & $B_p$  & -0.9710 & 0.8118 \\
        & $B_{\hat{p}}$ & 0.9052 & -0.8147 \\
        & $B_d$  & -0.9736 & 0.8267 \\
        & $B_{\hat{d}}$  & -0.1005  & 0.0989 \\
    \end{tabular}
    \caption{Correlation coefficients between weekly networks, NPI indices and fitted model parameters.
    }
    \label{tab:table2}
\end{table}

\section{Discussion}
% The changes to activity, reasons etc. effect of NPIs
In Finland the SARS-CoV-2 pandemic started in March 2020 having a visible effect on the activities of the population like their mobility, due to the government-imposed restrictions and self-imposed social distancing. This can be seen in Figure~\ref{fig:figure2} as a reduction in the overall activity in the bipartite networks and as a significant reduction in the average edge weight in the projected exposure networks during the pandemic.
Notably, we can see that higher level of activity does not explicitly mean higher exposure between the postal code areas as implied by the Spearman correlation between the two values being $\approx0.621$ with p-value $<0.01$.
Comparing the overall activity and edge weight to the NPIs in the form of an index, we see that the average edge weight has a higher correlation to the NPI stringency index than the activity has ($-0.824$ and $-0.668$). %($-0.8235$ and $-0.6682$).
High correlation to the NPI stringency index could be seen as high compliance to the restrictions as well as general awareness of the public, resulting in lesser exposure.
Moreover, the lesser amount of fluctuations, seasonal or other, in the average edge weight and the related fitted exponents of the models hint that the underlying behaviour remains stable during ''normal´´ pre-pandemic times but changes once the restrictions are imposed.
The changes in the network attributes such as assortativity depict the reorganization where instead of having more exposure to hubs of high weighted degree, the exposure is distributed more evenly across the postal code areas, as the assortativity shifts towards positive values. Similar trend can be seen in the weighted clustering coefficient, shifting to lower values when the pandemic sets on.

The proposed gravity and radiation type models are shown to be sufficient in estimating the exposure between postal code areas, with the gravity type model performing better than the radiation type model. As can be expected, the prediction error decreases once the gravity model is fitted with additional variables of population density, $\hat{p}$,  and public transport distance $\hat{d_{i,j}}$.
In addition to changing exponent weights, the error increases the more restricted human mobility is (see Fig.~\ref{fig:figureS2}). %~\ref{supp-fig:figure6}). 
While the error values can be seen to increase, also the $R^2$ values show an increase, which we suspect being caused by a higher number of extreme outliers. Fitting the models over the full duration of our data shows that the exponents (Fig.~\ref{fig:figure5}) correlate with the average edge weight as well as the NPI index. These changes in the exponents can be interpreted as shifting weight between distance and population size. Interestingly, in the extended gravity model the exponents for population size and population density show an anti-phase like behaviour. 
In both gravity type models, the weight for longer distances decreases (see Table \ref{tab:table1}), which we also confirmed manually from the empirical data. The distances larger than the nearest vicinity show a comprehensive reduction and the longest distances remain at the low level. Of this finding the government imposed restrictions and guidelines such as social distancing and remote work could explain the lower weight for long distance edges as people eligible for remote working would not be commuting. Temporary closures of restaurants and other services can also have a similar effect as the number of trips decreases~\cite{tully2021effect}.
The results in both projected networks and in the fitted model exponents showed a high correlation with the general NPI stringency index, and when fitting the series of exponents of the extended gravity model to the subcategories of restrictions we see that certain restrictions affect certain exponents more than others. The coarse model in itself provides an estimate where the extended gravity model's exponents would be given a certain set of restrictions. As mentioned before, using such exponents for the model yields a good estimate for the mean edge weight, but overestimates the network properties when compared to the observed networks.

% Technical points, limitations, design choices
% Data limitations
The mobile phone location based dataset used in this study has two limitations that could cause some inaccuracies.
Firstly, the data defines the home location of each person as the postal code area in which the person stayed longest before 9 am, i.e. the place the person woke up in the morning. This definition could misplace some groups of people like those staying in hotels, which could cause fluctuations in the number of activities per capita especially in the pre-pandemic times.
Secondly, the additional protection of individual privacy by filtering activities with less than five individuals from a postal code area creates inaccuracies in the projected networks, which is why in fitting the models we filter out the postal code areas with population less than 500. This filtering removes seven postal code areas, including three industrial areas with population sizes of $<200$, two sparsely populated areas with populations $<270$, a hospital and a military area. 

% Model limitations and assumptions
The design of this study considered an area-wise limited system of people living in 4 municipalities in capital yet most populated area of Finland.The limitation %of this closed system 
can cause inaccuracies with the radiation model in areas located in the outskirts of the investigated area, as the variable of people living within the radius~($r_{d_{i,j}}$) is not considering the populations outside of the system.
For the radiation model, the values for $r_{d_{i,j}}$ were also calculated such that the distribution of population was assumed to be even across the postal code area.
Another assumption was made for the population size data, as we only consider the numbers recorded in 2019. %due to conventionality. 
The changes between the beginning and the end of our dataset were not large enough to significantly change the results of the model. Due to this, we are not considering things such as influx of new inhabitants or outflow of people to other municipalities. 
A potential limitation and inaccuracy arises from calculation of distances between postal code areas. The assumption of having the geo-spatial center of each postal code area the point of calculation  disregards the population density and distribution within the area, which results in some level of error in distances especially between postal code areas sharing borders.
Also, in reality the distances between postal code areas are not the same as the calculated crow-flight distances, in which case a more realistic measure could be the travel time.
Considering the travel time between locations would most likely improve the model, but such data would add to the model complexity as the travel time can vary due to e.g. road closures or constructions and public transportation changes. %if certain roads are closed for instance due to construction or  the public transportation changes due to travel restrictions or addition or removal of stops.
Moreover, an average travel time weighted with the means of travel available to the local population could improve the results.

% Future, maybe to Conclusion
In the future we aim to conduct further
analysis on the empirical dataset in terms of the time series of exposure and the socioeconomic aspects of each postal code area. Also, a detailed analysis of the NPI effects is warranted~\cite{tully2021effect}.
Intuitively, the method of constructing a weighted projection network based on data of daily visits to a set of locations and using the resulting links as a proxy for contact could be used in modelling spreading processes between the postal code areas with compartmental SEIRS-type models~\cite{barrio2021model,snellman2022modeling}. 
%OMIT
%However, the network construction of this study is likely insufficient in itself to realistically capture the spreading of Covid-19 between the postal code areas as the type of exposure can vary depending on the grid size and the contents of therein (i.e. houses, businesses).
%Also, the definition of an activity in the empirical data does not explicitly mean that the two people stay in a location at the same time during the day.

\section{Conclusions}

Mobile phone based datasets have been shown to provide novel insights into human mobility and their social interactions.
In this paper we have investigated the use of aggregated mobile phone location data as a source for gaining insight into the exposure between populations living in different postal code areas using co-occurring visits to mobile phone grid locations as a proxy.
People visiting the same cellphone grid areas form a bipartite network, which we project using the Eq. \ref{eq:weight} resulting in a fully connected network with edge weights between 0 and 1.
The resulting empirical networks show the changes in exposure stemming from co-occurring activities between pre-pandemic and pandemic times shifting towards lower clustering, higher assortativity and lower weight for edges with longer distances between populations and to higher populated areas. These changes along with high correlation with the NPI-indices reflect the adherence to government imposed restrictions, remote work as well as self-imposed social distancing of aware individuals.
We showed that the weights of the exposure can be modelled using gravity and radiation type models with population variables and distances between the postal code areas. The simplistic models yield promising results at the level of average edge weights, but are found to be lacking in reproducing the assortativity and clustering of the empirical networks.
Lastly, we conducted a brief investigation on modelling exposure for the whole duration of our dataset using the different NPI sub-categories for calculating the exponents of our best performing gravity model, showing a varying degree of weight for the NPI sub-categories depending on the gravity model exponent.

\section*{Conflict of Interest Statement}

The authors declare that the research was conducted in the absence of any commercial or financial relationships that could be construed as a potential conflict of interest.

\section*{Author Contributions}

TT, KB and KK contributed to conception and design of the study. TT processed and analyzed the data and created the models. TT wrote the first draft of the manuscript and created the figures and tables. All authors contributed to manuscript revision, read, and approved the submitted version.

\section*{Funding}
T.T. acknowledges funding from Finnish Academy of Science and Letters Väisälä Foundation. This work was partly supported by NordForsk through the funding to The Network Dynamics of Ethnic Integration, project number 105147. K.B., and K.K. acknowledge support from EU HORIZON 2020 INFRAIA-2019-1 (SoBigData++) No. 871042.

%\section*{Acknowledgments}
%This is a short text to acknowledge the contributions of specific colleagues, institutions, or agencies that aided the efforts of the authors.

%\section*{Supplemental Data}
% \href{http://home.frontiersin.org/about/author-guidelines#SupplementaryMaterial}{Supplementary Material} should be uploaded separately on submission, if there are Supplementary Figures, please include the caption in the same file as the figure. LaTeX Supplementary Material templates can be found in the Frontiers LaTeX folder.

\section*{Data Availability Statement}
The data analyzed for this study is commercial and not publicly available.
%can be found in the [NAME OF REPOSITORY] [LINK].
% Please see the availability of data guidelines for more information, at https://www.frontiersin.org/about/author-guidelines#AvailabilityofData

\bibliographystyle{unsrtnat}
\bibliography{refs}

\clearpage

\section*{Supplementary Figures and Tables}

\begin{figure}[!h]
    \centering
    \includegraphics[width=1\textwidth]{a_performance_models_2019.pdf}
    \includegraphics[width=1\textwidth]{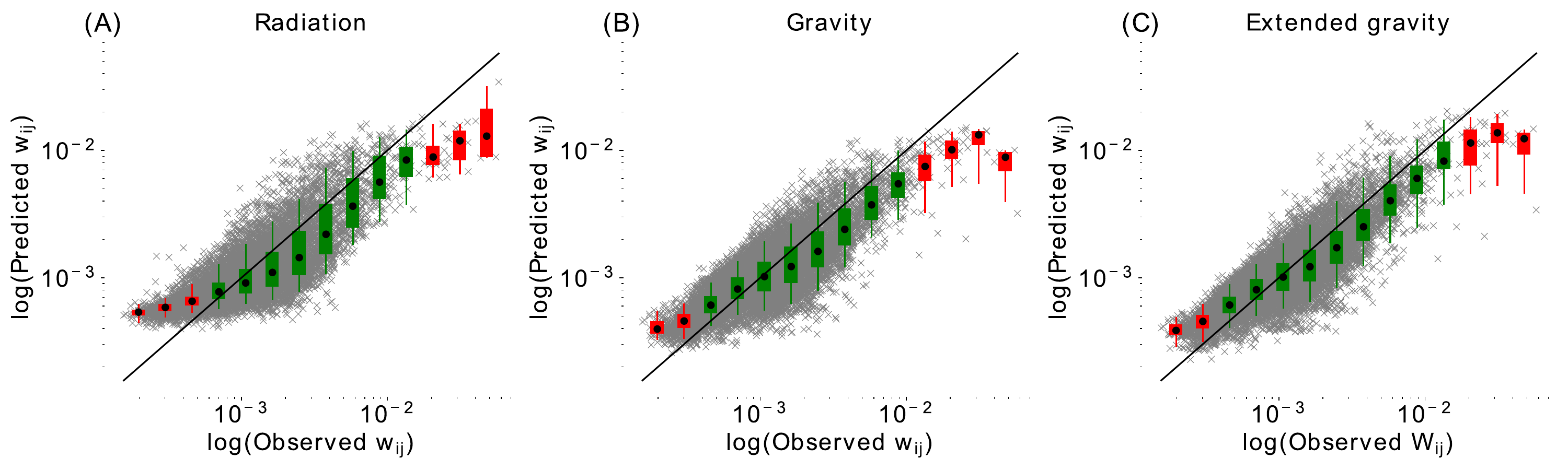}
    \includegraphics[width=1\textwidth]{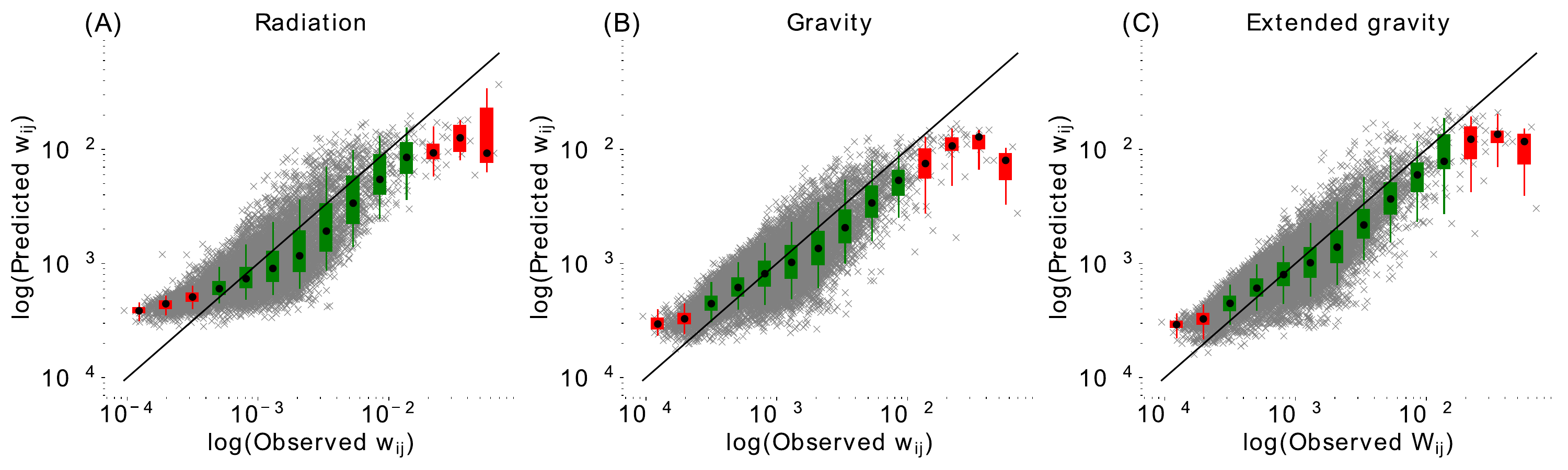}

    \caption{The predicted model values plotted against the observed empirical values for each yearly aggregated network 2019 (1st row), 2020 (2nd row) and 2021 (3rd row). Each point denotes an edge in the network and the straight line indicates the perfect fit. Points closer to the line have smaller error and points above the line are overestimates and vice versa. The points are visualized also in equally sized bins, depicting the mean (dot), the 25th percentile (box) and the 95th percentile (line connected to box). The colour of the box is green if the 95th percentile contains the perfect fit to the observed data (straight line).}\label{fig:figureS1}
\end{figure}

\begin{figure}[!h]
    \centering
    \includegraphics[width=1\textwidth]{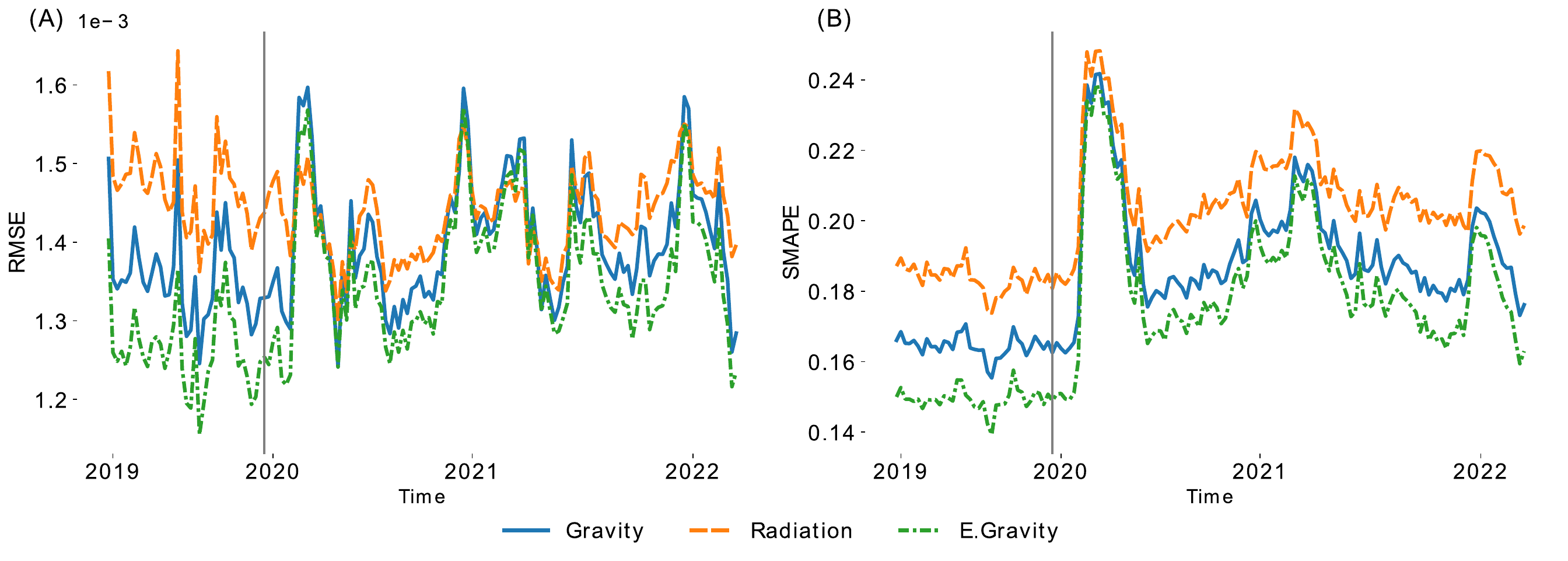}
    \caption{Error metrics for the gravity, radiation and extended gravity models over the duration of the data set, fitted on the weekly networks. (A) RMSE and (B) SMAPE. Lower values indicate a smaller error. The vertical line denotes the empty period in the data between 24th September 2019 and 31st January 2020.
    }\label{fig:figureS2}
\end{figure}

\begin{figure}[!h]
    \centering
    \includegraphics[width=0.95\textwidth]{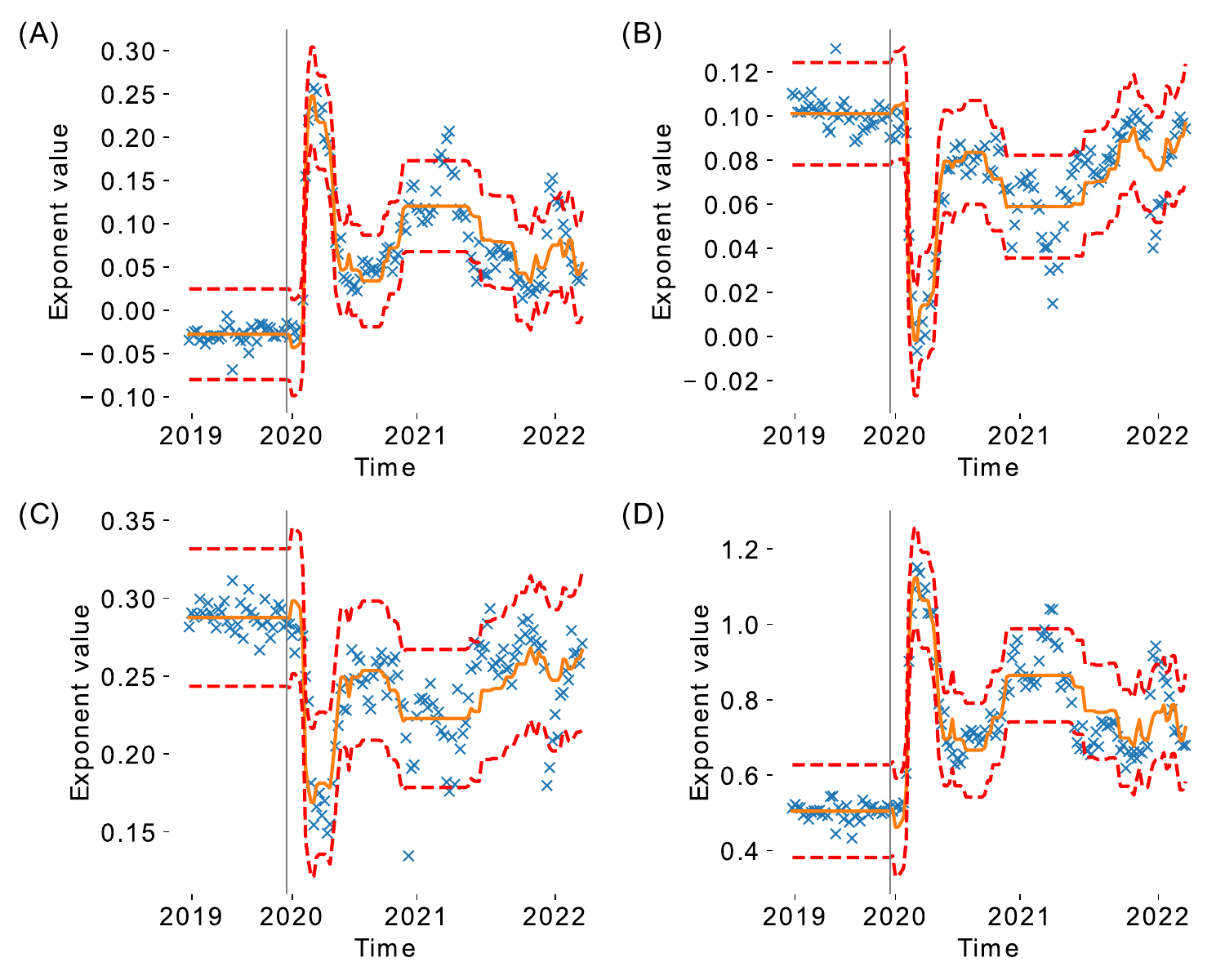}
    \caption{ The predicted exponents of the extended gravity model from an OLS regression fitted on the 8 indices of non-pharmaceutical interventions. The dashed lines depict the prediction's 95\% confidence interval. (A) is for population size exponent $B_p$, (B) is population density $B_{\hat{p}}$, (C) is for distance via public transport $B_{\hat{d}}$ and (D) is for distance $B_d$. 
    }\label{fig:figureS3}
\end{figure}

\def\arraystretch{1.5}
\begin{table}[!h]
    \centering
\begin{tabular}{l|c|c }
        Model result & $C$ & $B_p$ \\ \hline
        $R^2$ & 0.874 & 0.857 \\
        Constant & 0.0255$\pm$0.070, p=0.716 & -0.0283$\pm$0.003, p=0.000 \\
        H1 Public information campaigns & \textbf{0.0059$\pm$0.002, p=0.018} & -3.081e-05$\pm$0.000, p=0.782 \\
        C1M School closing & 0.0127$\pm$0.007, p=0.054 & \textbf{0.0008$\pm$0.000, p=0.005}  \\
        C2M Workplace closing & 0.0090$\pm$0.006, p=0.118 & 0.0003$\pm$0.000, p=0.313 \\
        C3M Cancel public events & \textbf{0.0075$\pm$0.003, p=0.025} & 0.0002$\pm$0.000, p=0.190 \\
        C4M Restrictions on gatherings &\textbf{ 0.0076$\pm$0.002, p= 0.000 }& \textbf{0.0003$\pm$7.57e-05, p=0.000} \\
        C5M Close public transport & 0.0083$\pm$0.007, p=0.265 & 0.0003$\pm$0.000, p=0.362 \\
        C6M Stay at home requirements & 0.0005$\pm$0.005, p=0.922 & 0.0002$\pm$0.000, p=0.469  \\
        C7M Restrictions on internal movement & 0.0087$\pm$0.003, p=0.008 &\textbf{ 0.0010$\pm$0.000, p=0.000} \\
        C8EV International travel controls & -0.0048$\pm$0.003, p=0.088 & -0.0002$\pm$0.000, p=0.196 \\
    
    \hline
    \hline    
    
        Model result  & $B_{\hat{p}}$ & $B_d$ \\ \hline
        $R^2$ &  0.847 & 0.599 \\
        Constant & 0.1008$\pm$0.002, p=0.000 & 0.2885$\pm$0.002, p=0.000 \\
        H1 Public information campaigns &  -3.057e-05$\pm$8.52e-05, p=0.720 & 0.0001$\pm$6.59e-05, p=0.055 \\
        C1M School closing & -0.0003$\pm$0.000, p=0.144 & \textbf{0.0004$\pm$0.000, p=0.018}   \\
        C2M Workplace closing  & \textbf{-0.0005$\pm$0.000, p=0.021}  &  \textbf{-0.0004$\pm$0.000, p=0.009}\\
        C3M Cancel public events & -0.0001$\pm$0.000, p=0.256 & 3.018e-05$\pm$8.86e-05, p=0.73 \\
        C4M Restrictions on gatherings & \textbf{-0.0002$\pm$5.8e-05, p=0.000 }& \textbf{0.0001$\pm$4.49e-05, p=0.008}  \\
        C5M Close public transport  & 0.0002$\pm$0.000, p=0.542 & \textbf{0.0005$\pm$0.000, p=0.009} \\
        C6M Stay at home requirements  & 1.805e-05$\pm$0.000, p=0.920 & 0.0001$\pm$0.000, p=0.312 \\
        C7M Restrictions on internal movement  & \textbf{-0.0007$\pm$0.000, p=0.000} & -0.0002$\pm$8.65e-05, p= 0.062 \\
        C8EV International travel controls  & 7.75e-05$\pm$9.74e-05, p=0.427 & 2.335e-05$\pm$7.53e-05, p=0.757 \\
     
     \hline
     \hline   
    
    \end{tabular}
    \begin{tabular}{l|c }
        Model result  & $B_{\hat{d}}$\\ \hline
        $R^2$ &  0.885\\
        Constant & 0.5044$\pm$0.011, p=0.000\\
        H1 Public information campaigns & 0.0006$\pm$0.000, p=0.129\\
        C1M School closing &\textbf{0.0028$\pm$0.001, p=0.006} \\
        C2M Workplace closing  & 0.0012$\pm$0.001, p=0.186\\
        C3M Cancel public events & \textbf{0.0012$\pm$0.001, p=0.023}\\
        C4M Restrictions on gatherings & \textbf{0.0012$\pm$0.000, p=0.000 }\\
        C5M Close public transport   & 0.0014$\pm$0.001, p=0.204 \\
        C6M Stay at home requirements  & 0.0003$\pm$0.001, p=0.735\\
        C7M Restrictions on internal movement  & \textbf{0.0024$\pm$0.000, p=0.000 }\\
        C8EV International travel controls  & \textbf{-0.0010$\pm$0.000, p=0.026}\\
    \hline 
        
    \end{tabular}
    \normalsize
    \caption{
    OLS regression results and fit for the NPI subcategories (rows) to extended gravity model's exponents (columns) over the the course of the dataset. The results are depicted as value of coefficient, standard error and the relate p-value of the fitting.
    Variables fitted with p-values$<.05$ are highlighted by bold text.
    }
    \label{tab:table3}
    
\end{table}

\end{document}